\documentclass[preprint, preprintnumbers ,amsmath,amssymb]{revtex4}
\usepackage{amssymb}
\usepackage{amsmath}
\usepackage{graphicx}
\usepackage{multirow}
\usepackage{makecell}

\begin{document}



\title{Hohenberg-Kohn Theorems in Electrostatic and Uniform Magnetostatic Fields}



\author{Xiao-Yin Pan}

\affiliation{Department of Physics, Ningbo University, Ningbo
315211, China}

\author{Viraht Sahni}

\affiliation{Brooklyn College and The Graduate School of the City
University of New York, 365 Fifth Avenue, New York, NY 10016}


\date{\today}

\begin{abstract}

The Hohenberg-Kohn (HK) theorems of bijectivity between the external
scalar potential and the gauge invariant nondegenerate ground state
density, and the consequent Euler variational principle for the
density, are proved for arbitrary electrostatic field and the
constraint of fixed electron number.  The HK theorems are
generalized for spinless electrons to the added presence of an
external uniform magnetostatic field by introducing the new
constraint of fixed canonical orbital angular momentum.  Thereby a
bijective relationship between the external scalar and vector
potentials, and the gauge invariant nondegenerate ground state
density and physical current density, is proved.  A corresponding
Euler variational principle in terms of these densities is also
developed.  These theorems are further generalized to electrons with
spin by imposing the added constraint of fixed canonical orbital and
spin angular momentum.  The proofs differ from the original HK
proof, and explicitly account for the many-to-one relationship
between the potentials and the nondegenerate ground state wave
function.

\end{abstract}

\pacs{}

\maketitle



\section{Introduction}

The Hohenberg-Kohn (HK) theorems \cite{1} constitute a fundamental
advance in quantum mechanics.  As a consequence they have furthered
our understanding of the electronic structure of matter: atoms,
molecules, solids, clusters, surfaces, lower dimensional electronic
systems such as heterostructures, quantum dots, graphene, etc.
Matter, according to HK, is described as a system of  $N$ electrons
in an external electrostatic field ${\boldsymbol{\cal{E}}}
({\bf{r}}) = - {\boldsymbol{\nabla}} v ({\bf{r}})$.  The first HK
theorem defines the concept of a \emph{basic variable} of quantum
mechanics. Knowledge of this gauge invariant property  -- the
nondegenerate ground state density $\rho ({\bf{r}})$ -- is of
two-fold significance: \emph{(a)} It determines the Schr\"{o}dinger
theory wave functions $\Psi$ of the system, both ground and excited
state; \emph{(b)} As the wave function $\Psi$ is now proved to be a
functional of the basic variable, it constitutes together with the
second HK theorem  -- the energy variational principle for arbitrary
variations of the density -- the basis of theories of electronic
structure such as of Hohenberg-Kohn \cite{1}, Kohn-Sham \cite{2},
and quantal density functional theory \cite{3,4}. The theorems are
valid for arbitrary confining potential $v ({\bf{r}})$ and electron
number $N$, but are derived \cite{5} for the constraint of
\emph{fixed} $N$. In this paper we generalize the HK theorems for
spinless electrons to the added presence of an external
\emph{uniform} magnetostatic field.  As the presence of the magnetic
field constitutes a new degree of freedom, we introduce the further
natural constraint of \emph{fixed} canonical orbital angular
momentum. Thereby we prove that the basic variables in quantum
mechanics in a uniform magnetic field are the gauge invariant
nondegenerate ground state density $\rho ({\bf{r}})$ and physical
current density ${\bf{j}} ({\bf{r}})$.  These theorems are then
further generalized to electrons with spin by imposing the
constraints of \emph{fixed} canonical orbital and spin angular
momentum.

The generalization is motivated by the considerable recent interest
in yrast states which are states of lowest energy for fixed angular
momentum. These states have been studied experimentally and
theoretically for both bosons and fermions, e.g. rotating trapped
Bose-Einstein condensates \cite{6}, and harmonically trapped
electrons in the presence of a uniform perpendicular magnetic field
\cite{7}.  The theorems derived are applicable to all
experimentation with a uniform magnetic field such as the
magneto-caloric  effect \cite{8}, the Zeeman effect, cyclotron
resonance, magnetoresistance, the de-Haas-van Alphen effect, the
Hall effect, the quantum Hall effect, the Meissner effect, nuclear
magnetic resonance, etc.

The manner by which a basic variable is so defined is via the proof
of the first HK theorem for $v$-representable densities.  To explain
this, and to contrast the present proofs with the HK proof, we first
briefly describe the HK arguments.  The HK theorems are proved for a
nondegenerate ground state. Particularizing to electrons without any
loss of generality, the Hamiltonian $\hat{H}$ in atomic units
(charge of electron $-e; |e| = \hbar = m = 1)$ is $\hat{H} =
\frac{1}{2} \sum_{k} p^{2}_{k} + \frac{1}{2} \sum'_{k, \ell}
1/|{\bf{r}}_{k} - {\bf{r}}_{\ell}| + \sum_{k} v ({\bf{r}}_{k})$,
where the terms correspond to the kinetic $\hat{T}$ (with momentum
$\hat{\bf{p}}_{k} = - i {\boldsymbol{\nabla}}_{{\bf{r}}_{k}}$), the
electron-interaction $\hat{W}$, and external potential $\hat{V}$
operators, respectively. The Schr\"{o}dinger equation is $\hat{H}
({\bf{R}}) \Psi ({\bf{X}}) = E \Psi ({\bf{X}})$, where $\Psi
({\bf{X}}), E$ are the eigenfunctions and eigenenergies, with
${\bf{R}} = {\bf{r}}_{1}, \ldots, {\bf{r}}_{N}$; ${\bf{X}} =
{\bf{x}}_{1}, \ldots, {\bf{x}}_{N}$; ${\bf{x}} = {\bf{r}} \sigma$
being the spatial and spin coordinates of the electron. The energy
$E$ is the expectation $E = <\Psi ({\bf{X}}) | \hat{H} ({\bf{R}}) |
\Psi ({\bf{X}}) >$. In the first HK theorem it is initially proved
(Map C) that there is a one-to-one relationship between the external
potential $v ({\bf{r}})$ and the nondegenerate ground-state wave
function $\Psi ({\bf{X}})$. \emph{Employing this relationship}, it
is then proved (Map D) that there is a one-to-one relationship
between the wave function $\Psi ({\bf{X}})$ and the corresponding
nondegenerate ground state density $\rho ({\bf{r}})$.   Thus,
knowledge of $\rho ({\bf{r}})$ determines $v ({\bf{r}})$ to within a
constant.  Since for a fixed electron number $N$, the kinetic
$\hat{T}$ and electron-interaction potential $\hat{W}$ energy
operators are known, so is the system Hamiltonian.  Solution of the
corresponding Schr\"{o}dinger equation then leads to the wave
functions $\Psi$  of the system.  \emph{It is the one-to-one
relationship between the external potential and the gauge invariant
density that defines the latter as a basic variable}.  As the wave
function $\Psi$, and hence energy $E_{v} [ \rho]$ are functionals of
the density $\rho ({\bf{r}})$, the variational Euler equation for
the density with \emph{fixed} $v ({\bf{r}})$ follows subject to the
constraint of \emph{known} electron number $N$ (see Table 1). (The
lowest nondegenerate \cite{9,10} excited state density $\rho^{e}
({\bf{r}})$ of a given symmetry different from that of the ground
state is also a basic variable.)

In the added presence of an external magnetostatic field ${\bf{B}}
({\bf{r}}) = \nabla \times {\bf{A}} ({\bf{r}})$, where ${\bf{A}}
({\bf{r}})$ is the vector potential, the Hamiltonian when the
interaction of the field is only with the orbital angular momentum
is
\begin{equation}
\hat{H} = \frac{1}{2} \sum_{k} \bigg[ \hat{\bf{p}}_{k} + \frac{1}{c}
{\bf{A}} ({\bf{r}}_{k}) \bigg]^{2} + \hat{W} + \hat{V}.
\end{equation}
When the interaction of the magnetic field is with both the orbital
and spin angular momentum, the Hamiltonian is
\begin{equation}
\hat{H} = \frac{1}{2} \sum_{k} \bigg[ \hat{\bf{p}}_{k} + \frac{1}{c}
{\bf{A}} ({\bf{r}}_{k}) \bigg]^{2} + \hat{W} + \hat{V} + \frac{1}{c}
\sum_{k} {\bf{B}} ({\bf{r}}_{k}) \cdot {\bf{s}}_{k},
\end{equation}
where ${\bf{s}}$ is the electron spin angular momentum vector.  In
deriving the Hamiltonians of Eqs. (1) and (2), we have hewed to the
philosophy \cite{11} that the only `fundamental' interactions are
those that can be generated by the substitution $\hat{\bf{p}}
\rightarrow \hat{\bf{p}} + \frac{1}{c} {\bf{A}}$.  (This then
defines the physical momentum operator in the presence of a magnetic
field, and thereby the physical current density ${\bf{j}}
({\bf{r}})$.)  In non-relativistic quantum mechanics, the
Hamiltonian of Eq. (2) is derived \cite{11} by Schr\"{o}dinger-Pauli
theory for spin $\frac{1}{2}$ particles via the kinetic energy
operator $\frac{1}{2} \boldsymbol{\sigma} \cdot ({\bf{p}} +
{\bf{A}}) \boldsymbol{\sigma} \cdot ({\bf{p}} + {\bf{A}})$, where
$\boldsymbol{\sigma}$ is the Pauli matrix, and ${\bf{s}} =
\frac{1}{2} \boldsymbol{\sigma}$. The spin magnetic moment generated
in this way has the correct gyromagnetic ratio $g = 2$.

It would appear that one could prove a one-to-one relationship
between the gauge invariant properties $\{ \rho ({\bf{r}}), {\bf{j}}
({\bf{r}}) \}$ and the external potentials $\{ v ({\bf{r}}),
{\bf{A}} ({\bf{r}}) \}$ along the lines of the HK proof. However, no
such proof is possible as the relationship between the external
potentials $\{ v ({\bf{r}}), {\bf{A}} ({\bf{r}}) \}$ and the
non-degenerate ground state wave function $\Psi ({\bf{X}})$ can be
\emph{many-to-one} \cite{12} and even \emph{infinite-to-one}
\cite{13}.  Hence, in these cases, there is no equivalent of Map C,
and therefore the original HK path is not possible. The proof that
$\{ \rho ({\bf{r}}), {\bf{j}} ({\bf{r}}) \}$ are the basic variables
must then differ from the original HK proof. Furthermore, the proof
must account for the many-to-one relationship between $\{ v
({\bf{r}}), {\bf{A}} ({\bf{r}}) \}$ and $\Psi ({\bf{X}})$.

In the literature \cite{2,12,14}, the proofs of what properties
constitute the basic variables are not rigorous in the HK sense of
the one-to-one relationship between the basic variables and the
external potentials $\{ v, {\bf{A}} \}$.  Further, they do not
account for the many-to-one relationship between $\{ v, {\bf{A}} \}$
and $\Psi$. Additionally, the system angular momentum is not
considered. The choice of the basic variables is arrived at solely
on the basis of a Map D-type proof between these assumed properties
and the nondegenerate ground state $\Psi$, thereby the claim that
$\Psi$ is a functional of these properties. In these proofs, the
existence of a bijective Map C is implicitly assumed, \cite{15,16}
(see also last reference of 12). For example, in spin-DFT
\cite{2,12,14} for which the Hamiltonian is that of Eq. (2) with the
field component of the momentum absent, the basic variables are
assumed to be $\{ \rho ({\bf{r}}), {\bf{m}} ({\bf{r}}) \}$, where
${\bf{m}} ({\bf{r}})$ is the magnetization density. In current-DFT
\cite{14}, corresponding to the Hamiltonian of Eq. (1), the basic
variables are assumed to be $\rho ({\bf{r}})$ and the gauge variant
paramagnetic current density ${\bf{j}}_{p} ({\bf{r}})$.  For the
Hamiltonian of Eq. (2), the basic variables are assumed to be $\{
\rho ({\bf{r}}), {\bf{m}} ({\bf{r}}), {\bf{j}}_{p} ({\bf{r}}) \}$ or
$\{ \rho ({\bf{r}}), {\bf{m}} ({\bf{r}}), {\bf{j}}_{p} ({\bf{r}}),
{\bf{j}}_{p, {\bf{m}}} ({\bf{r}}) \}$ where ${\bf{j}}_{p, {\bf{m}}}
({\bf{r}})$ are the gauge variant paramagnetic currents of each
component of the magnetization density. Subsequently, a Map D proof
is provided. Additionally, with the basic variables now assumed
known, a Percus-Levy-Lieb (PLL)-type proof \cite{17,18} can then be
formulated \cite{19}. More recently, we gave a derivation
\cite{15,20} which purported to prove that $\{ \rho ({\bf{r}}),
{\bf{j}} ({\bf{r}}) \}$ were the basic variables but the proof was
in error \cite{21}. Subsequently, we proved \cite{22} for the
Hamiltonian of Eq. (1) that for the significant subset of systems
\cite{13,23} for which the ground state wave function $\Psi$ is
real, the basic variables are $\{ \rho ({\bf{r}}), {\bf{j}}
({\bf{r}}) \}$.  Our proof of bijectivity between $\{ \rho
({\bf{r}}), {\bf{j}} ({\bf{r}}) \}$ and $\{ v ({\bf{r}}), {\bf{A}}
({\bf{r}}) \}$ explicitly accounts for the many-to-one $\{ v
({\bf{r}}), {\bf{A}} ({\bf{r}}) \}$ to $\Psi$ relationship.  This
proof then constitutes a special case of the more general proof for
$\Psi$ complex presented in this work.

Here we extend the HK theorems to systems of electrons in external
electrostatic ${\boldsymbol{\cal{E}}} ({\bf{r}}) = -
{\boldsymbol{\nabla}}  v ({\bf{r}})$ and magnetostatic ${\bf{B}}
({\bf{r}}) = \nabla \times {\bf{A}} ({\bf{r}})$ fields with known
electron number $N$ and angular momentum ${\bf{J}}$. The proofs are
for a uniform magnetostatic field, and for Hamiltonians in which the
interaction of the magnetic field is \emph{(i)} solely with the
orbital angular momentum $({\bf{J}} = {\bf{L}})$, and \emph{(ii)}
with both the orbital and spin angular momentum $({\bf{J}} =
[{\bf{L}}$ and ${\bf{S}}])$.  We prove, in the \emph{rigorous} HK
sense, that for \emph{fixed} $N$ and ${\bf{J}}$ the basic variables
are the gauge invariant nondegenerate ground state density $\rho
({\bf{r}})$ and \emph{physical} current density ${\bf{j}}
({\bf{r}})$. In other words, knowledge of $\{ \rho ({\bf{r}}),
{\bf{j}} ({\bf{r}}) \}$ determines the potentials $\{v ({\bf{r}}),
{\bf{A}} ({\bf{r}}) \}$ to within a constant and the gradient of a
scalar function, respectively. Hence, with the Hamiltonians known,
solution of the respective Schr\"{o}dinger and Schr\"{o}dinger-Pauli
equations lead to the wave functions of each system. The proof is
for $(v, {\bf{A}})$-representable $\{ \rho ({\bf{r}}), {\bf{j}}
({\bf{r}}) \}$. The extension to the Percus-Levy-Lieb (PLL)
\cite{17,18} constrained-search path for $N$-representable and
degenerate states readily follows. As the wave function $\Psi$ is a
functional of $\{ \rho ({\bf{r}}), {\bf{j}} ({\bf{r}}) \}$, theories
of electronic structure based on $\{ \rho ({\bf{r}}), {\bf{j}}
({\bf{r}}) \}$ as the basic variables can then be formulated.

\section{Proof of Generalized Hohenberg-Kohn Theorems}

To accentuate the role of the density $\rho ({\bf{r}})$ and physical
current density ${\bf{j}} ({\bf{r}})$, we rewrite the Hamiltonians
of Eqs. (1) and (2) in terms of operators representative of these
gauge invariant properties.  The Hamiltonians can then be written,
respectively, as
\begin{equation}
\hat{H} = \hat{T} + \hat{W} + \hat{V}_{A},
\end{equation}
and
\begin{equation}
\hat{H} = \hat{T} +  \hat{W} + \hat{V}_{A} - \int \hat{\bf{m}}
({\bf{r}}) \cdot {\bf{B}} ({\bf{r}}) d {\bf{r}},
\end{equation}
where the total external potential operator $\hat{V}_{A}$ is
\begin{equation}
\hat{V}_{A} = \hat{V} + \frac{1}{c} \int \hat{\bf{j}} ({\bf{r}})
\cdot {\bf{A}} ({\bf{r}}) d {\bf{r}} - \frac{1}{2c^{2}} \int
\hat{\rho} ({\bf{r}}) A^{2} ({\bf{r}}) d {\bf{r}},
\end{equation}
and the corresponding energy expectations $E = <\Psi ({\bf{X}}) |
\hat{H} | \Psi ({\bf{X}}) >$ as
\begin{equation}
E = T + E_{ee} + V_{A},
\end{equation}
and
\begin{equation}
E = T + E_{ee} + V_{A} - \int {\bf{m}} ({\bf{r}}) \cdot {\bf{B}}
({\bf{r}}) d {\bf{r}},
\end{equation}
where the total external potential energy $V_{A}$ is
\begin{equation}
V_{A} = ~ < \Psi ({\bf{X}}) | \hat{V}_{A} | \Psi ({\bf{X}}) = \int
\rho ({\bf{r}}) v ({\bf{r}}) d {\bf{r}} + \frac{1}{c} \int {\bf{j}}
({\bf{r}}) \cdot {\bf{A}} ({\bf{r}}) d {\bf{r}} - \frac{1}{2c^{2}}
\int \rho ({\bf{r}}) A^{2} ({\bf{r}}) d {\bf{r}},
\end{equation}
and where $T$ and $E_{ee}$ are the kinetic and electron-interaction
energy expectations.  In the above equations, the physical current
density ${\bf{j}} ({\bf{r}})$ is defined in terms of the physical
momentum operator $(\hat{\bf{p}} + \frac{1}{c} {\bf{A}})$ as
\begin{equation}
{\bf{j}} ({\bf{r}}) = N \Re \sum_{\sigma} \int \Psi^{\star}
({\bf{r}} \sigma, {\bf{X}}^{N-1}) \bigg(\hat{\bf{p}} + \frac{1}{c}
{\bf{A}} ({\bf{r}})\bigg) \Psi ({\bf{r}} \sigma, {\bf{X}}^{N-1}) d
{\bf{X}}^{N-1},
\end{equation}
or equivalently as the expectation of the current density operator
$\hat{\bf{j}} ({\bf{r}})$:
\begin{equation}
{\bf{j}} ({\bf{r}}) = < \Psi ({\bf{X}}) | \hat{\bf{j}} ({\bf{r}}) |
\Psi ({\bf{X}}) >
\end{equation}
where
\begin{equation}
\hat{\bf{j}} ({\bf{r}}) = \hat{\bf{j}}_{p} ({\bf{r}}) +
\hat{\bf{j}}_{d} ({\bf{r}}),
\end{equation}
with the paramagnetic $\hat{\bf{j}}_{p} ({\bf{r}})$ and diamagnetic
$\hat{\bf{j}}_{d} ({\bf{r}})$ operator components defined,
respectively, as
\begin{equation}
\hat{\bf{j}}_{p} ({\bf{r}}) = \frac{1}{2} \sum_{k} \big[
\hat{\bf{p}}_{k} \delta ({\bf{r}}_{k} - {\bf{r}}) + \delta
({\bf{r}}_{k} - {\bf{r}}) \hat{\bf{p}}_{k} \big],
\end{equation}
and
\begin{equation}
\hat{\bf{j}}_{d} ({\bf{r}}) = \hat{\rho} ({\bf{r}}) {\bf{A}}
({\bf{r}})/c,
\end{equation}
with the density operator $\hat{\rho} ({\bf{r}})$ being
\begin{equation}
\hat{\rho} ({\bf{r}}) = \sum_{k} \delta ({\bf{r}}_{k} - {\bf{r}}).
\end{equation}
The magnetization density ${\bf{m}} ({\bf{r}})$ is the expectation
\begin{equation}
{\bf{m}} ({\bf{r}}) = <\Psi ({\bf{X}}) | \hat{\bf{m}} ({\bf{r}}) |
\Psi ({\bf{X}}) >,
\end{equation}
with the local magnetization density operator $\hat{\bf{m}}
({\bf{r}})$ defined as
\begin{equation}
\hat{\bf{m}} ({\bf{r}}) = - \frac{1}{c} \sum_{k} {\bf{s}}_{k} \delta
({\bf{r}}_{k} - {\bf{r}}).
\end{equation}
(The current density operator $\hat{\bf{j}} ({\bf{r}})$ can also be
defined in terms of the Hamiltonian $\hat{H}$ as $\hat{\bf{j}}
({\bf{r}}) = c \partial \hat{H}/\partial {\bf{A}}$.  This confirms
that for both the Hamiltonians of Eqs. (3) and (4), the physical
current density is the orbital current density.)

We first present the proof of bijectivity between $\{ \rho
({\bf{r}}), {\bf{j}} ({\bf{r}}) \}$ and $\{ v ({\bf{r}}), {\bf{A}}
({\bf{r}}) \}$ for spinless electrons corresponding to the
Hamiltonian of Eq. (1) or (3) for fixed electron number $N$ and
angular momentum ${\bf{L}}$. The proof is by \emph{reductio ad
absurdum}. Let us consider two different physical systems $\{ v,
{\bf{A}} \}$ and $\{ v', {\bf{A}}' \}$ that generate different
nondegenerate ground state wave functions $\Psi$ and $\Psi'$.  We
assume the gauges of the unprimed and primed systems to be the same.
Let us further assume that these systems lead to the \emph{same}
nondegenerate ground state $\{ \rho ({\bf{r}}), {\bf{j}} ({\bf{r}})
\}$.  We prove this cannot be the case.  From the variational
principle for the energy for a nondegenerate ground state, one
obtains the inequality
\begin{equation}
E = ~ <\Psi | \hat{H} | \Psi > ~ < ~ <\Psi' | \hat{H} | \Psi' >.
\end{equation}
Now
\begin{eqnarray}
<\Psi' | \hat{H} | \Psi'> = < \Psi' | \hat{T} &+& \hat{W} + \hat{V}'
+ \frac{1}{c} \int \hat{\bf{j}}' ({\bf{r}}) \cdot {\bf{A}}'
({\bf{r}}) d
{\bf{r}} \nonumber \\
&-& \frac{1}{2c^{2}} \int \hat{\rho} ({\bf{r}}) A'^{2} ({\bf{r}}) d
{\bf{r}} | \Psi' > + < \Psi' | \hat{V} - \hat{V}' | \Psi'> \nonumber
\\ &+& \frac{1}{c} < \Psi' |  \int [ \hat{\bf{j}} ({\bf{r}}) \cdot {\bf{A}}
({\bf{r}}) - \hat{\bf{j}}' ({\bf{r}}) \cdot {\bf{A}}' ({\bf{r}})] d
{\bf{r}}| \Psi' > \nonumber \\
&-& \frac{1}{2c^{2}} < \Psi' | \int \hat{\rho} ({\bf{r}}) [A^{2}
({\bf{r}}) - A'^{2} ({\bf{r}})] d {\bf{r}} | \Psi' >.
\end{eqnarray}
Employing the above assumptions, and following the same steps as in
\cite{22}, one obtains the inequality
\begin{equation}
E + E' < E + E' + \int \big[ {\bf{j}}'_{p} ({\bf{r}}) - {\bf{j}}_{p}
({\bf{r}}) \big] \cdot \big[ {\bf{A}} ({\bf{r}}) - {\bf{A}}'
({\bf{r}}) \big] d {\bf{r}},
\end{equation}
where $E' = <\Psi' | \hat{H}' | \Psi' >$.

As the majority of the experimental and consequent theoretical work
is performed for uniform magnetic fields, our proof too is for such
fields.

Consider next the third term on the right hand side of Eq. (19).
With ${\bf{B}} ({\bf{r}}) = B \hat{\bf{i}}_{z}$, ${\bf{B}}'
({\bf{r}}) = B' \hat{\bf{i}}_{z}$, and the symmetric gauge ${\bf{A}}
({\bf{r}}) = \frac{1}{2} {\bf{B}} \times {\bf{r}}$, ${\bf{A}}'
({\bf{r}}) = \frac{1}{2} {\bf{B}}' \times {\bf{r}}$, this term may
be written as
\begin{equation}
I = \frac{1}{2} \Delta {\bf{B}} \cdot \int {\bf{r}} \times \bigg[
{\bf{j}}'_{p} - {\bf{j}}_{p} ({\bf{r}}) \bigg] d {\bf{r}},
\end{equation}
where $\Delta {\bf{B}} = (B - B') \hat{\bf{i}}_{z}$.  First consider
the integral
\begin{eqnarray}
I_{1} &=& \int {\bf{r}} \times {\bf{j}}_{p} ({\bf{r}})  d {\bf{r}} \\
&=& - \frac{i} {2} \sum_{k} \int d {\bf{X}} \int d {\bf{r}}
\Psi^{\star} ({\bf{X}}) \big[ {\bf{r}} \times
{\boldsymbol{\nabla}}_{{\bf{r}}_{k}} \delta ({\bf{r - r}}_{k}) +
\delta ({\bf{r - r}}_{k}) {\bf{r}} \times
{\boldsymbol{\nabla}}_{{\bf{r}}_{k}} \big] \Psi ({\bf{X}}).
\end{eqnarray}
Next consider the second integral of $I_{1}$ of Eq. (22):
\begin{eqnarray}
I_{12} &=& \frac{1} {2} \int d {\bf{X}} \Psi^{\star} ({\bf{X}})
\big( \sum_{k} {\bf{r}}_{k}
\times \hat{\bf{p}}_{k} \big) \Psi ({\bf{X}}) \\
&=& \frac{1} {2} \int d {\bf{X}} \Psi^{\star} ({\bf{X}}) \sum_{k}
\hat{\bf{L}}_{k} \Psi ({\bf{X}}) = \frac{1}{2} {\bf{L}},
\end{eqnarray}
where $\hat{\bf{L}}_{k} = {\bf{r}}_{k} \times \hat{\bf{p}}_{k}$ is
the canonical orbital angular momentum operator, with $\hat{\bf{p}}$
the canonical momentum operator $(\hat{\bf{p}} =
\hat{\bf{p}}_{kinetic} + \hat{\bf{p}}_{field} = m {\bf{v}} +
\frac{q}{c} {\bf{A}})$, and ${\bf{L}}$ the total canonical orbital
angular momentum defined by Eq. (24).  Note that the canonical
angular momentum is gauge variant.

The first integral of $I_{1}$ of Eq. (22) is
\begin{equation}
I_{11} = - \frac{i}{2} \sum_{k} \int d {\bf{X}} \int d {\bf{r}}
\Psi^{\star} ({\bf{X}}) \epsilon_{\alpha \beta \gamma}
\frac{\partial} {\partial r_{k \gamma}} \big(r_{\beta} \delta
({\bf{r - r}}_{k})  \Psi ({\bf{X}}) \big).
\end{equation}
On integrating the inner integral by parts and dropping the surface
term, one obtains
\begin{eqnarray}
I_{11} &=& - \frac{i}{2} \sum_{k} \int d {\bf{X}} \big [ -
\epsilon_{\alpha \beta \gamma} \int d {\bf{r}} \frac{\partial
\Psi^{\star} ({\bf{X}})} {\partial r_{k \gamma}} r_{\beta} \delta
({\bf{r -
r}}_{k}) \Psi ({\bf{X}}) \big] \\
&=& - \frac{i}{2} \sum_{k} \int d {\bf{X}} \big [ - \epsilon_{\alpha
\beta \gamma} \frac{\partial \Psi^{\star} ({\bf{X}})} {\partial r_{k
\gamma}} r_{k \beta} \Psi ({\bf{X}}) \big].
\end{eqnarray}
On integrating by parts again, one obtains
\begin{eqnarray}
I_{11} &=& - \frac{i}{2} \sum_{k} \epsilon_{\alpha \beta \gamma}
\int d {\bf{X}} \Psi^{\star} ({\bf{X}})  \frac{\partial} {\partial
r_{k \gamma}}
\big(r_{k \beta} \Psi ({\bf{X}}) \big) \\
&=& - \frac{i}{2} \sum_{k} \int d {\bf{X}} \Psi^{\star} ({\bf{X}}) ~
({\bf{r}}_{k} \times {\boldsymbol{\nabla}}_{{\bf{r}}_{k}}) \Psi
({\bf{X}}) = \frac{1}{2} {\bf{L}}
\end{eqnarray}
Hence, the integral $I$ of Eq. (20) is
\begin{equation}
I = \frac{1}{2} \Delta {\bf{B}} \cdot ({\bf{L}}' - {\bf{L}}).
\end{equation}
If one imposes the condition that the total canonical orbital
angular momentum is \emph{fixed} so that ${\bf{L}} = {\bf{L}}'$,
then the integral $I$ vanishes so that the third term on the right
hand side of Eq. (19) vanishes.

For states with fixed orbital angular momentum ${\bf{L}}$, Eq. (19)
then reduces to the contradiction
\begin{equation}
E + E' < E + E'.
\end{equation}
What this means is that the original assumption that $\Psi$ and
$\Psi'$ differ is erroneous, and that there can exist a $\{ v,
{\bf{A}} \}$ and a $\{ v', {\bf{A}}' \}$ with the same nondegenerate
ground state wave function. The fact that $\Psi=\Psi'$ means that
$\rho ({\bf{r}}) |_{\Psi} = \rho' ({\bf{r}}) |_{\Psi'}$.  However,
the corresponding physical current densities are not the same:
${\bf{j}} ({\bf{r}}) |_{\Psi} \neq {\bf{j}}' ({\bf{r}}) |_{\Psi'}$,
because ${\bf{j}}_{d} ({\bf{r}})|_{\Psi} \neq
{\bf{j}}'_{d}|_{\Psi'}$ if one hews with the original assumption
that ${\bf{A}} ({\bf{r}})$ is different from ${\bf{A}}' ({\bf{r}})$.
This proves that the assumption that there exists a different $\{
v', {\bf{A}}' \}$ (with the same $N$ and ${\bf{L}}$) that leads to
the same $\{ \rho, {\bf{j}} \}$ as that due to $\{ v, {\bf{A}} \}$
is incorrect. This step takes into account the fact that there could
exist many $\{ v, {\bf{A}} \}$ that lead to the same nondegenerate
ground state $\Psi$. Hence, there exists only one $\{ v, {\bf{A}}
\}$ for fixed $N$ and ${\bf{L}}$ that leads to a nondegenerate
ground state $\{ \rho, {\bf{j}} \}$. The one-to-one relationship
between $\{ \rho, {\bf{j}} \}$ and $\{ v, {\bf{A}} \}$ is therefore
proved for the case when the interaction of the magnetic field is
solely with the orbital angular momentum.

With $ \{ \rho ({\bf{r}}), {\bf{j}} ({\bf{r}}) \}$ as the basic
variables, the wave function $\Psi$ is a functional of these
properties. By a density and physical current density
\emph{preserving} unitary transformation \cite{4,15,24} it can be
shown that the wave function must also be a functional of a gauge
function $\alpha ({\bf{R}})$. This ensures that the wave function
when written as a functional: $\Psi = \Psi [\rho, {\bf{j}}, \alpha]$
is gauge variant. However, as the physical system remains unchanged
for different gauge functions, the choice of vanishing gauge
function is valid.

As the ground state energy is a functional of the basic variables:
$E = E_{v, {\bf{A}}} [\rho, {\bf{j}}]$, a variational principle for
$E_{v, {\bf{A}}} [\rho, {\bf{j}}]$ exists for arbitrary variations
of $(v, {\bf{A}})$-representable densities $ \{ \rho ({\bf{r}}),
{\bf{j}} ({\bf{r}}) \}$. The corresponding Euler equations for $\rho
({\bf{r}})$ and ${\bf{j}} ({\bf{r}})$ follow, and these must be
solved self-consistently with the constraints $\int \rho ({\bf{r}})
d {\bf{r}} = N$, $\int {\bf{r}} \times ({\bf{j}} ({\bf{r}}) -
\frac{1}{c} \rho ({\bf{r}}) {\bf{A}} ({\bf{r}}) d {\bf{r}} =
{\bf{L}}$ and ${\boldsymbol{\nabla}} \cdot {\bf{j}} ({\bf{r}}) = 0$.
Implicit in this variational principle, as in all such energy
variational principles, \emph{is that the external potentials remain
fixed throughout the variation}.  (See Table I.)

We next consider electrons with spin corresponding to the
Hamiltonian of Eq. (2) or (4). In this case, with the same
assumptions made regarding the two different physical systems $\{ v,
{\bf{A}}; \psi \}$ and $\{ v', {\bf{A}}'; \psi' \}$ leading to the
same $\{ \rho ({\bf{r}}), {\bf{j}} ({\bf{r}})$ as before, the
inequality of Eq. (19) is replaced by
\begin{eqnarray}
E + E' < E + E' &+& \int \big[ {\bf{j}}'_{p} ({\bf{r}}) -
{\bf{j}}_{p} ({\bf{r}}) \big] \cdot \big[ {\bf{A}} ({\bf{r}}) -
{\bf{A}}' ({\bf{r}}) \big] d {\bf{r}} \nonumber \\
&-& \int \big[ {\bf{m}}' ({\bf{r}}) - {\bf{m}} ({\bf{r}}) \big]
\cdot \big[ {\bf{B}} ({\bf{r}}) - {\bf{B}}' ({\bf{r}}) \big] d
{\bf{r}}.
\end{eqnarray}
The third term on the right hand side vanishes if one imposes the
constraint that the orbital angular momentum ${\bf{L}}$ of the
unprimed and primed systems are the same.  Hence, next consider the
last term of Eq. (32).  For a uniform magnetic field with ${\bf{B}}
({\bf{r}}) = B \hat{\bf{i}}_{z}$ and ${\bf{B}}' ({\bf{r}}) = B
\hat{\bf{i}}_{z}$, we have
\begin{equation}
\int {\bf{m}} ({\bf{r}}) \cdot {\bf{B}} ({\bf{r}}) d {\bf{r}} = B
\int m_{z} ({\bf{r}}) d {\bf{r}},
\end{equation}
where \cite{19}
\begin{equation}
m_{z} ({\bf{r}}) = - \frac{1}{2c} \big[ \rho_{\alpha} ({\bf{r}}) -
\rho_{\beta} ({\bf{r}}) \big],
\end{equation}
with $\rho_{\alpha} ({\bf{r}}), \rho_{\beta} ({\bf{r}})$ being the
spin-up and spin-down spin densities.  The last term of the
inequality is then
\begin{equation}
\int \big[ {\bf{m}}' ({\bf{r}}) - {\bf{m}} ({\bf{r}}) \big] \cdot
\Delta {\bf{B}} ({\bf{r}}) d {\bf{r}} = - \frac{1} {2c} \Delta B
\int \big[ \{ \rho'_{\alpha} ({\bf{r}}) - \rho'_{\beta} ({\bf{r}})
\} - \{ \rho_{\alpha} ({\bf{r}}) - \rho_{\beta} ({\bf{r}}) \} \big]
d {\bf{r}},
\end{equation}
with $\Delta B = B - B'$.  If the $z$-component of the total spin
angular momentum $S_{z}$ for the unprimed and primed systems are the
same, the corresponding spin densities are the same.  The last term
of Eq. (35) thus vanishes leading once again to the contradiction $E
+ E' < E + E'$.  More generally, the magnetization densities
${\bf{m}} ({\bf{r}})$ and ${\bf{m}}' ({\bf{r}})$ are the same if the
total spin angular momentum ${\bf{S}}$ are the same.  Hence, once
again, the bijective relationship between the nondegenerate ground
state densities $\{ \rho ({\bf{r}}), {\bf{j}} ({\bf{r}}) \}$ and the
potentials $\{ v ({\bf{r}}), {\bf{A}} ({\bf{r}}) \}$ is proved
provided one imposes the constraint that the total orbital
${\bf{L}}$ and spin ${\bf{S}}$ angular momentum are fixed.

This may be seen in a different manner by accentuating the role of
the spin angular momentum.  With the $z$-component of the total spin
${\bf{S}}$ being $S_{z} = \sum_{k} s_{z,k}$, the density $m_{z}
({\bf{r}})$ may be written as
\begin{equation}
m_{z} ({\bf{r}}) = - \frac{1}{cN} \sum_{\sigma} S_{z} \gamma
({\bf{r}} \sigma, {\bf{r}} \sigma),
\end{equation}
with $\gamma ({\bf{x x}}') = N \int \Psi^{\star} ({\bf{r}} \sigma,
{\bf{X}}^{N-1}) \Psi ({\bf{r}}' \sigma', {\bf{X}}^{N-1}) d
{\bf{X}}^{N-1}$, the density matrix.  Since in the primed system,
the spin vectors are different, i.e. some ${\bf{s}}'_{k}$, we have
\begin{eqnarray}
\int \big[ {\bf{m}}' ({\bf{r}}) - {\bf{m}} ({\bf{r}}) \big] \cdot
\Delta {\bf{B}} ({\bf{r}}) d {\bf{r}} &=& \Delta B \int \big[ m'_{z}
({\bf{r}}) - m_{z} ({\bf{r}}) \big] d {\bf{r}} \\
&=& \frac{\Delta B} {c N} \sum_{\sigma} \int \big[ S'_{z} \gamma'
({\bf{r}} \sigma, {\bf{r}} \sigma) - S_{z} \gamma ({\bf{r}} \sigma,
{\bf{r}} \sigma) \big] d {\bf{r}}.
\end{eqnarray}
Employing the original assumption that the diagonal matrix elements
$\gamma ({\bf{r}} \sigma, {\bf{r}} \sigma)$ of the density matrix
$\gamma ({\bf{x x}}')$ are the same for the unprimed and primed
systems we have the right hand side of Eq. (38) to be
\begin{equation}
\frac{\Delta B} {c N} \sum_{\sigma} \int \big[ S'_{z} - S_{z} \big]
\gamma ({\bf{r}} \sigma, {\bf{r}} \sigma) = 0
\end{equation}
provided $S'_{z} = S_{z}$.

In the above proofs for the Hamiltonians of Eqs (3) and (4), the
definition of the current density ${\bf{j}} ({\bf{r}})$ employed is
that of Eq. (10).  However, for finite systems, the Hamiltonian of
Eq. (4) can also be written as \cite{25}
\begin{eqnarray}
\hat{H} = \hat{T} + \hat{W} + \hat{V} + \frac{1}{c} \int
\hat{\bf{j}}_{p} ({\bf{r}}) \cdot {\bf{A}} ({\bf{r}}) d {\bf{r}} &+&
\frac{1} {2c^{2}} \int \hat{\rho} ({\bf{r}}) A^{2} ({\bf{r}}) d
{\bf{r}} \nonumber \\
&+& \frac{1}{c} \int \hat{\bf{j}}_{m} ({\bf{r}}) \cdot {\bf{A}}
({\bf{r}}) d {\bf{r}},
\end{eqnarray}
where the magnetization current density operator $\hat{\bf{j}}_{m}
({\bf{r}})$ is defined as
\begin{equation}
\hat{\bf{j}}_{m} ({\bf{r}}) = - c {\boldsymbol{\nabla}} \times
{\bf{m}} ({\bf{r}}).
\end{equation}
Hence the physical current density ${\bf{j}} ({\bf{r}})$ may also be
defined as \cite{25}
\begin{equation}
{\bf{j}} ({\bf{r}}) = c \frac{\partial \hat{H}} {\partial {\bf{A}}
({\bf{r}})} = {\bf{j}}_{p} ({\bf{r}}) + {\bf{j}}_{d} ({\bf{r}}) +
{\bf{j}}_{m} ({\bf{r}}),
\end{equation}
the sum of the paramagnetic, diamagnetic, and magnetization current
densities. Even for this definition of the physical current density
${\bf{j}} ({\bf{r}})$, the proof of bijectivity between $\{ \rho,
{\bf{j}} \}$ and $\{ v, {\bf{A}} \}$ is valid provided the angular
momentum ${\bf{L}}$ and ${\bf{S}}$ are fixed.  For spin-compensated
systems, the magnetization current density ${\bf{j}}_{m} ({\bf{r}})$
vanishes.

\section{Concluding Remarks}

In conclusion, we have generalized the HK theorems to the added
presence of a uniform magnetic field.  We have considered the cases
of the interaction of the magnetic field with the orbital angular
momentum as well as when the interaction is with both the orbital
and spin angular momentum.  In this work we have proved a one-to-one
relationship between the external potentials $\{ v ({\bf{r}}),
{\bf{A}} ({\bf{r}}) \}$ and the nondegenerate ground state densities
$\{ \rho ({\bf{r}}), {\bf{j}} ({\bf{r}}) \}$.  The proof differs
from that of the original HK theorem, and explicitly accounts for
the many-to-one relationship between the potentials $\{ v
({\bf{r}}), {\bf{A}} ({\bf{r}}) \}$ and the nondegenerate ground
state wave function $\Psi$.  To account for the presence of the
magnetic field, which constitutes an added degree of freedom, one
must then impose a further constraint beyond that of fixed electron
number $N$ as in the original HK theorems.  For the Hamiltonian
corresponding to spinless electrons, the added constraint is that of
fixed canonical orbital angular momentum ${\bf{L}}$.  For that
corresponding to electrons with spin, the constraints imposed are
those of fixed canonical orbital ${\bf{L}}$ and spin ${\bf{S}}$
angular momentum. (The gauge employed for the canonical angular
momentum ${\bf{L}}$ can be chosen to be the same as that employed
for the Hamiltonian.) It is the further constraint on the angular
momentum that makes a rigorous HK-type proof of bijectivity between
the gauge invariant basic variables and the external scalar and
vector potentials possible. Additionally, the HK-type proofs are
possible because the Hamiltonians considered are rigorously derived
from the tenets of nonrelativistic quantum mechanics.

With the knowledge that the basic variables are $\{ \rho ({\bf{r}}),
{\bf{j}} ({\bf{r}}) \}$, a variational principle for the energy
functional $E_{v, {\bf{A}}} [\rho, {\bf{j}}]$ for arbitrary
variations of $(v, {\bf{A}})$-representable densities $\{ \rho
({\bf{r}}), {\bf{j}} ({\bf{r}}) \}$ is then developed for each
Hamiltonian considered. The constraints on the corresponding Euler
equations are those of fixed electron number and angular momentum,
and the satisfaction of the equation of continuity.

Again, knowing what the basic variables are, it is possible to map
the interacting system defined by the Hamiltonians of Eqs. (1) and
(2) to one of noninteracting fermions with the same $\rho
({\bf{r}}), {\bf{j}} ({\bf{r}})$, and ${\bf{J}}$.  Such a mapping
has been derived within QDFT \cite{26}.  The theory has been applied
to map an interacting system \cite{13} of two electrons in a
magnetic field and a harmonic trap $v ({\bf{r}}) = \frac{1}{2}
\omega_{0} r^{2}$ for which the ground state wave function is $\Psi
({\bf{r}}_{1}, {\bf{r}}_{2}) = C (1 + r_{12}) e^{- \frac{1}{2}
(r^{2}_{1} r^{2}_{2})}$, where $r_{12} = | {\bf{r}}_{1} -
{\bf{r}}_{2}|$ and $C^{2} = 1/\pi^{2} (3 + \sqrt{2 \pi})$, to one of
noninteracting fermions with the same $ \{ \rho ({\bf{r}}), {\bf{j}}
({\bf{r}}) \}$.  This example corresponds to the special case of
zero angular momentum.  However, the QDFT mapping for finite angular
momentum is straight forward. For other recent work see
\cite{27,28}.  The conclusions in the latter are based on the
assumption of existence of a HK theorem but one without the
requirement of the constraint on the angular momentum.

X.-P. was supported by the National Natural Sciences Foundation of
China, Grant No. 11275100, and the K.C. Wong Magna Foundation of
Ningbo University.  The work of V.S. was supported in part by the
Research Foundation of the City University of New York.

\newpage

\begin{table}
\begin{tabular}{|c|c|c|}
\hline
\makecell {\textbf{Theory}} &  {~\textbf{Hohenberg-Kohn  DFT}~} & {\textbf{Generalized HK DFT}} \\
\hline \makecell{Parameters characterizing \\ ground state } &
\makecell{Electron Number $N$} &
\makecell{Electron Number $N$  \\ Angular momentum ${\bf{L}}$ }\\
\hline \makecell{Relationship between \\ potentials and wave
function  } & \makecell{One-to-one between\\ $v ({\bf{r}})$ and
$\Psi$  } & \makecell{Many-to-one between \\ $\{ v ({\bf{r}}),
{\bf{A}} ({\bf{r}}) \}$ and $\Psi$} \\
\hline \makecell{Properties characterizing \\ ground state} &
\makecell{Electron density $\rho ({\bf{r}})$} & \makecell{Electron
density $\rho ({\bf{r}})$
\\ Physical current density ${\bf{j}} ({\bf{r}})$}\\
\hline \makecell{Bijectivity theorem} &
\makecell{For fixed $N$ \\
$\rho ({\bf{r}}) \leftrightarrow v ({\bf{r}})$}  &
\makecell{For fixed $N$ and ${\bf{L}}$ \\
$\{ \rho ({\bf{r}}), {\bf{j}} ({\bf{r}}) \} \leftrightarrow \{v ({\bf{r}}), {\bf{A}} ({\bf{r}}) \}$} \\
\hline \makecell{Wave function \\ and Energy functionals } &
\makecell{$\Psi = \Psi [\rho, \alpha]$ \\ For fixed $v: E=E_{v}
[\rho]$ } & \makecell {$\Psi = \Psi [\rho, {\bf{j}}, \alpha]$ \\ For
fixed $\{ v, {\bf{A}} \} : E=E_{v, {\bf{A}}} [\rho, {\bf{j}} ]$ } \\
\hline \makecell{Euler equations \\ and constraints} &
\makecell{Variational principle for  \\
fixed $v$ and known $N$: \\ $\frac{\delta E_{v} [\rho]} {\delta
\rho} = 0$ \\ $\int \rho ({\bf{r}}) d {\bf{r}} = N$} &
\makecell{Variational principle for  \\
fixed $\{v, {\bf{A}} \}$ and known $N, {\bf{L}}$: \\ $\frac{\delta
E_{v, {\bf{A}}} [\rho, {\bf{j}}]} {\delta \rho}\bigg |_{\bf{j}} = 0$
~~~~ $\frac{\delta E_{v, {\bf{A}}} [\rho, {\bf{j}}]} {\delta
{\bf{j}}} \bigg|_{\rho} = 0$ \\ $\int \rho ({\bf{r}}) d {\bf{r}} =
N$
\\ {$\int {\bf{r}} \times ({\bf{j}} ({\bf{r}}) - \frac{1}{c} \rho ({\bf{r}})
{\bf{A}} ({\bf{r}})) d {\bf{r}} = {\bf{L}}$}  \\
${\boldsymbol{\nabla}} \cdot {\bf{j}} ({\bf{r}}) = 0$ } \\
\hline
\end{tabular}
\caption{Comparison of Hohenberg-Kohn and Generalized Hohenberg-Kohn
theories.}
\end{table}

\end{document}